\documentclass[aps,prb,preprint,groupedaddress,floatfix]{revtex4}
\usepackage{graphicx,latexsym}

\begin{document}

%\preprint{}

\title{Reentrant Adhesion Behavior in Nanocluster Deposition}


\author{A. Awasthi and S. C. Hendy}
\affiliation{MacDiarmid Institute for Advanced Materials
and Nanotechnology, Industrial Research Ltd, Lower Hutt 5040, New Zealand}
\author{P. Zoontjens}
\affiliation{MacDiarmid Institute for Advanced Materials
and Nanotechnology, School of Chemical and Physical Sciences, Victoria University of Wellington,
Wellington 6140, New Zealand}
\author{S. A. Brown}
\affiliation{MacDiarmid Institute for Advanced Materials and Nanotechnology, Department of Physics and Astronomy,
University of Canterbury, Christchurch 8140, New Zealand}

%\thanks{}

%\altaffiliation{MacDiarmid Institute of Advanced Materials and Nanotechnology,
%School of Chemical and Physical Sciences, Victoria University of Wellington, PO Box 600, Wellington, 
%New Zealand}

\date{\today}

\begin{abstract}
We simulate the collision of atomic clusters with a weakly attractive surface using molecular dynamics in a regime between soft-landing and fragmentation, where the cluster undergoes large deformation but remains intact. As a function of incident kinetic energy, we find a transition from adhesion to reflection at low kinetic energies. We also identify a second adhesive regime at intermediate kinetic energies, where strong deformation of the cluster leads to an increase in contact area and adhesive energy.    
\end{abstract}

%\pacs{}
%\keywords{}

\maketitle


%\subsection{}
%\subsubsection{}

\begin{figure}
\resizebox{\columnwidth}{!}{\includegraphics{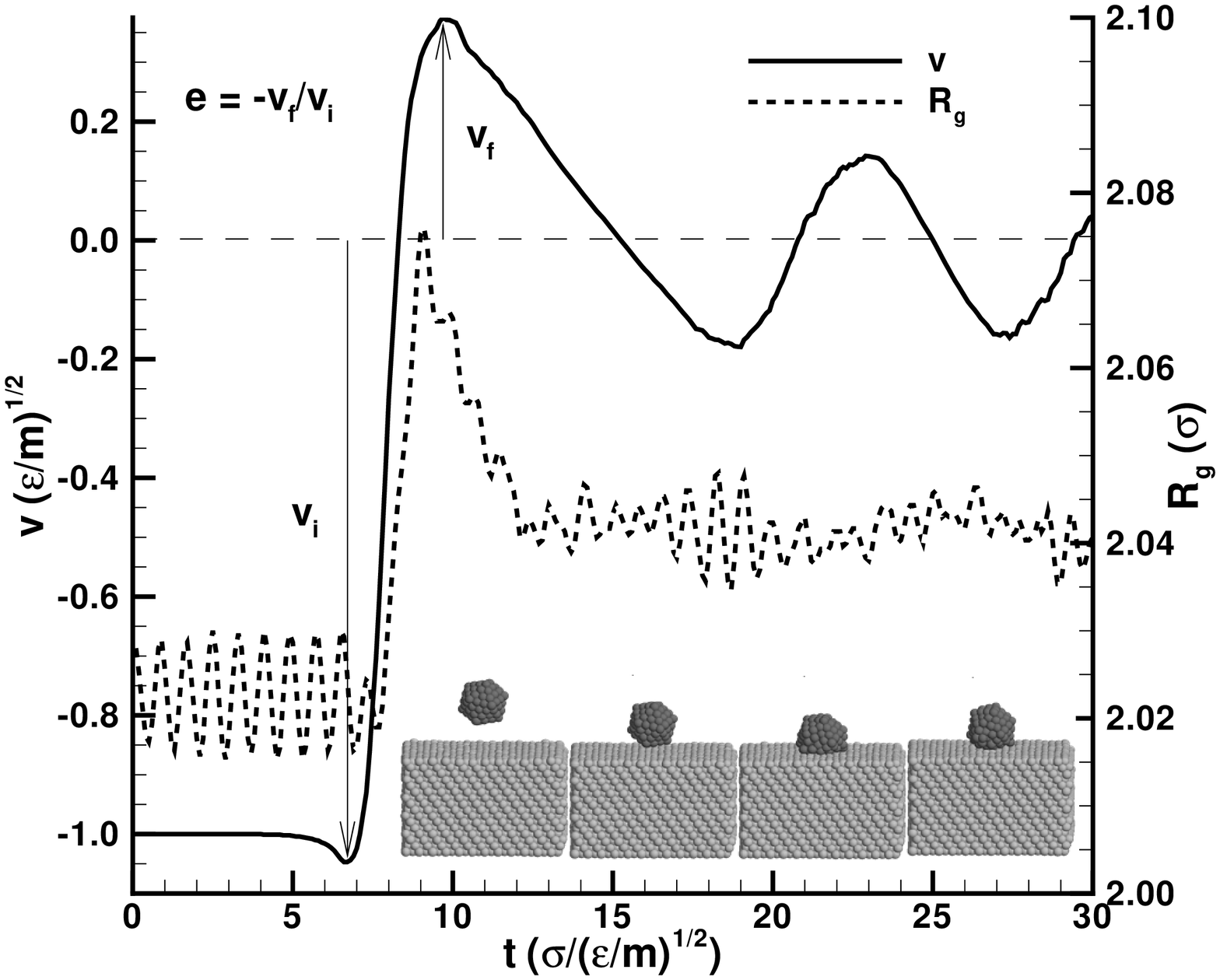}}
\caption{\label{DefnRest} The evolution of the velocity and radius of gyration of the 147-atom cluster is shown over the duration of the collision illustrated in a series of snap-shots. The cluster has an initial velocity of $1.0 \, (\varepsilon/m)^{1/2}$. In this case the cluster is captured by the surface and the velocity oscillates about zero after the collision.}
\end{figure}

The quantitative discourse on inelastic collisions between solid bodies stretches back at least as far as Newton's treatment in the Principia. Indeed,  the inelasticity of collisions in flowing granular materials is frequently still described by Newton's coefficient of restitution, $e$, which is the ratio of the reflected to incident velocity (see Fig.~\ref{DefnRest}). Although the coefficient of restitution is often regarded as a material constant, generally it is found to depend on both the incident velocity and the adhesion between the solid objects \cite{Johnson87}. Indeed, a full description of inelastic collisions is still a matter of on-going interest for the study of granular materials \cite{brilliantov04}. Nonetheless while inelastic collisions have been studied extensively for collisions of micro- or milliscale particles \cite{Hutchings77,Dave97,Wu03}, much less is known about the collision of nanoscale objects such as atomic clusters.  

The collision of clusters with surfaces has been studied by many groups in the context of cluster deposition (for a recent review see Ref. \cite{harbich00}). However, although the possibility of cluster reflection from hard surfaces has been considered (see the ``phase" diagrams in Refs.~\cite{harbich00,hsieh92}), early studies focused on implantation \cite{yamamura94}, fragmentation \cite{chatelet92, pettersson93} or soft-landing \cite{cheng94}. In the so-called soft-landing regime, the incident kinetic energy is generally insufficient to overcome the adhesion between the surface and the cluster, resulting in collisions that always lead to adhesion. At high kinetic energies, one enters the fragmentation regime where the cluster fragments upon impact into atoms or several large components \cite{chatelet92}, or the implantation regime where the cluster buries itself in the surface \cite{Palmer02}. Only recently has an intermediate regime been identified where clusters were observed to undergo a transition from adhesion to reflection \cite{partridge04}. This transition was exploited to assemble nanowires. 

Here we report on a detailed molecular dynamics study of the collisions of clusters, with kinetic energies that lie between the soft-landing and fragmentation regimes, on surfaces with a weak attraction to the cluster. The collisions span a range from weakly to strongly inelastic, and can lead to substantial deformation of the cluster. We have considered collisions of 147, 309 and 561-atom Mackay icosahedra with a fcc (111)-terminated slab. The interaction between cluster atoms and between surface atoms was modeled using a single Lennard-Jones potential (in what follows $\sigma$ denotes the Lennard-Jones radius, $\varepsilon$ denotes the Lennard-Jones bond energy and $m$ denotes the atomic mass). However the adhesion energy between the cluster and the surface was controlled by reducing the attractive $(\sigma/r)^6$ term between cluster atoms and surface atoms \cite{barrat99} by a factor $c$ (which was varied between 0.3 and 0.7). For instance when a 147-atom cluster was placed on the surface, it relaxed to form a contact angle of $120^\circ$ at $c=0.7$ and $100^\circ$ at $c=0.5$. In Ref~\cite{partridge04} a contact angle of $120^\circ$ for Sb clusters on SiO$_2$ was estimated. Here we report results for $c=0.35$ and only comment on trends for other values of $c$. We have also used a large cut-off distance of $6 \, \sigma$ in the potential, which ensured that the adhesion energy between the cluster and the surface was well converged and enabled us to probe a collision regime where the kinetic energy and adhesion energy were delicately balanced.

In each simulation, a central region of atoms in the slab follow Newtonian dynamics, an exterior `shell' follow Langevin dynamics \cite{Langevin86} at a temperature $T$ and a lower layer is fixed. In the Langevin region the friction parameter was varied linearly from 0 at the Langevin-Newtonian interface to 2 at the Langevin exterior to minimize reflections of elastic waves from boundaries \cite{Pomeroy02}. This shell of Langevin atoms regulates the temperature of the central Newtonian region and absorbs energy from the cluster impact allowing us to use a much smaller slab than would otherwise be possible. The size of the Newtonian and Langevin regions was chosen by studying a series of slabs of increasing depth and breadth until the energetics of test collisions were observed to have converged. The slab finally chosen was 11.7 $\sigma$ by 11.3 $\sigma$ in plane and 10.3 $\sigma$ deep, comprising 5846 Langevin atoms plus 1344 Newtonian atoms.  

\begin{figure}
\resizebox{\columnwidth}{!}{\includegraphics{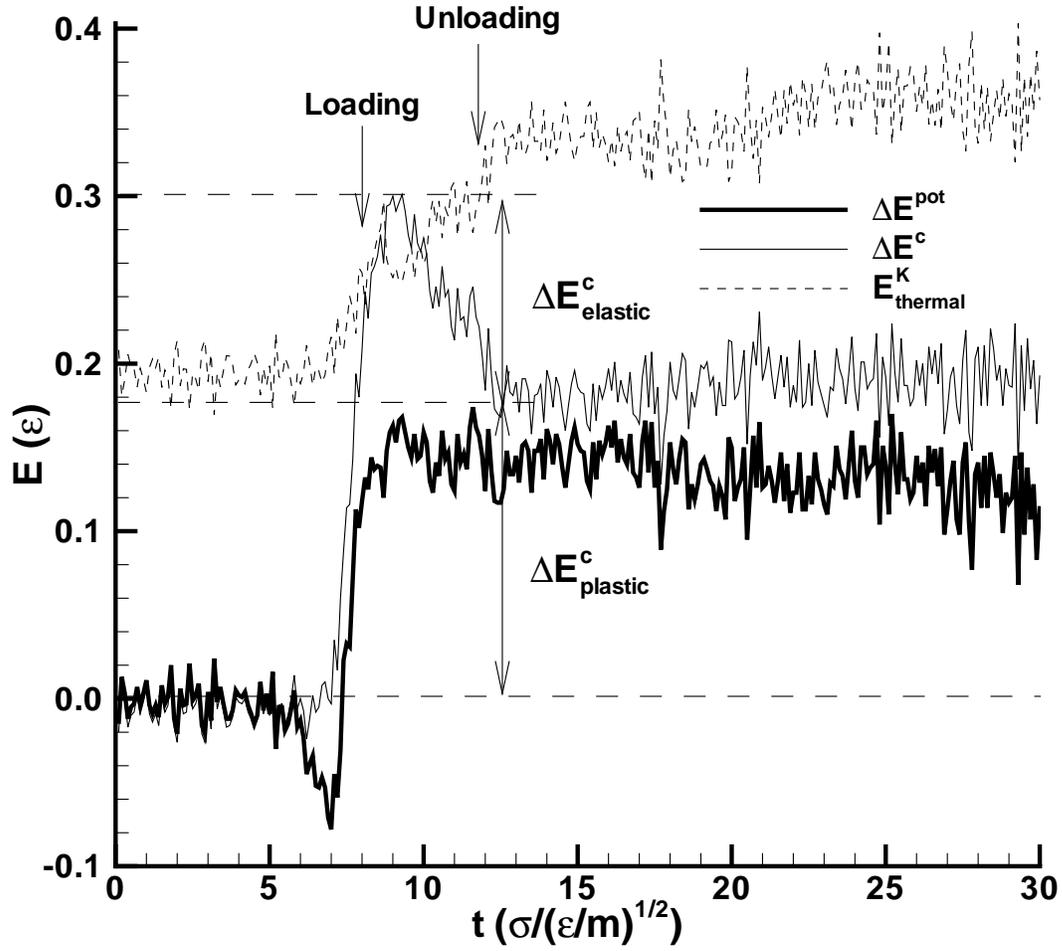}}
\caption{\label{StickFigure} The change in cluster potential energy (both the cluster internal energy $\Delta E^c$ and cluster total potential energy $\Delta E^{pot}$) and the thermal kinetic energy ($E^K_{thermal}$) are shown during the collision in Fig.~\ref{DefnRest}.} 
\end{figure}

\begin{figure}
\resizebox{\columnwidth}{!}{\includegraphics{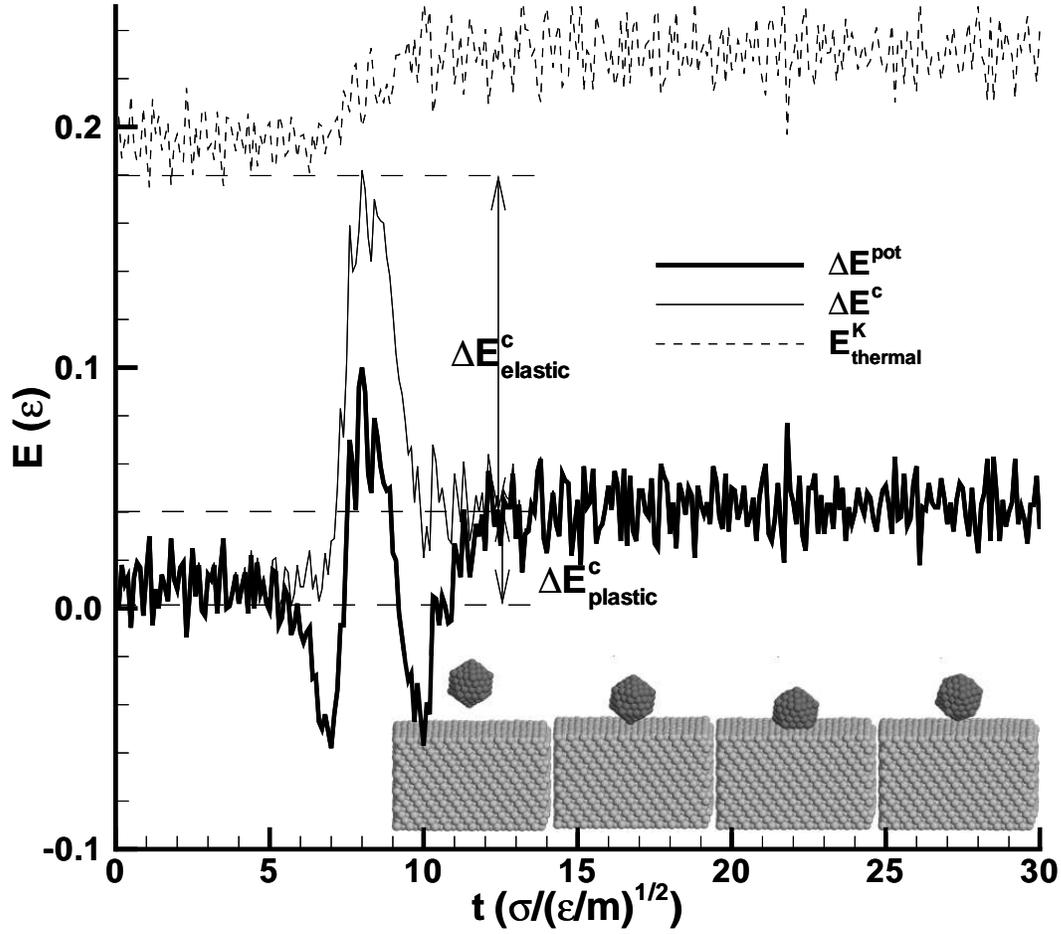}}
\caption{\label{BounceFigure} The change in the cluster potential energy per atom ($\Delta E^c$), the change in total potential energy per atom ($\Delta E^{pot}$), the thermal kinetic energy per atom $E^K_{thermal}$ and the center of mass velocity are shown as a function of time in a collision that results in reflection. The cluster is a 147-atom icosahedron and its initial velocity is $1.0 \,(\varepsilon/m)^{1/2}$.} 
\end{figure}

Figures~\ref{DefnRest}-\ref{BounceFigure} show the results from two collisions involving a 147-atom icosahedral cluster at an initial velocity $v_0 = 1.0 \,(\varepsilon/m)^{1/2}$ with $c=0.35$. Prior to collision, the cluster was equilibrated at a temperature of 0.13 $\varepsilon/k_B$ for $10^4$ time-steps and the surface was equilibrated at 0.2 $\varepsilon/k_B$ (the cluster was equilibrated at a lower temperature to avoid evaporation). Figure~\ref{DefnRest} shows the velocity and deformation of a 147-atom cluster for a collision that results in adhesion. The potential energy of the cluster during this collision is shown in Figure~\ref{StickFigure}. Figure \ref{BounceFigure} shows the potential energy for an event that leads to a reflection. Note that in the collision that leads to adhesion, the cluster is edge-on to the surface. However in the collision that results in reflection, the cluster is orientated with a vertex closest to the surface. This difference in orientation leads to substantially different deformation of the cluster and dissipation of kinetic energy upon impact. 

Figure~\ref{StickFigure} shows both the change in internal energy of the cluster per atom $\Delta E^c = E^c - E^c_{i}$  (where $E^c_{i}$ is the potential energy per atom of the isolated cluster) and the change in total potential energy per atom of the cluster $\Delta E^{pot} = \Delta E^c + \Delta E^{cs}$ (where $E^{cs}$ is the cluster-surface interaction energy per cluster atom). At impact $\Delta E^c$ rises sharply as the cluster undergoes both reversible (elastic) and irreversible (plastic) deformation. Note that the loading and unloading phases of the collision are evident in both the change in the radius of gyration $R_g$ and the two-step change in the thermal kinetic energy $E^K_{thermal}$. At the end of the unloading phase, we can see that while some of the internal energy of the cluster has been recovered ($\Delta E^c_{elastic}$), most of the change in energy is irreversible ($\Delta E^c_{plastic}$). In the case of reflection (Fig.~\ref{BounceFigure}) the cluster undergoes substantially less deformation than in the previous case, and the majority of the change in cluster internal energy is elastic.  
		
It is evident that the outcome of a collision depends strongly on orientation and thus cluster structure. However in what follows, we have averaged out this effect by conducting 50 trials for each set of collision parameters, with cluster orientation varied randomly between trials. We can then discuss the probability of bouncing at a given velocity and for a prescribed surface and cluster. 

\begin{figure}
\resizebox{\columnwidth}{!}{\includegraphics{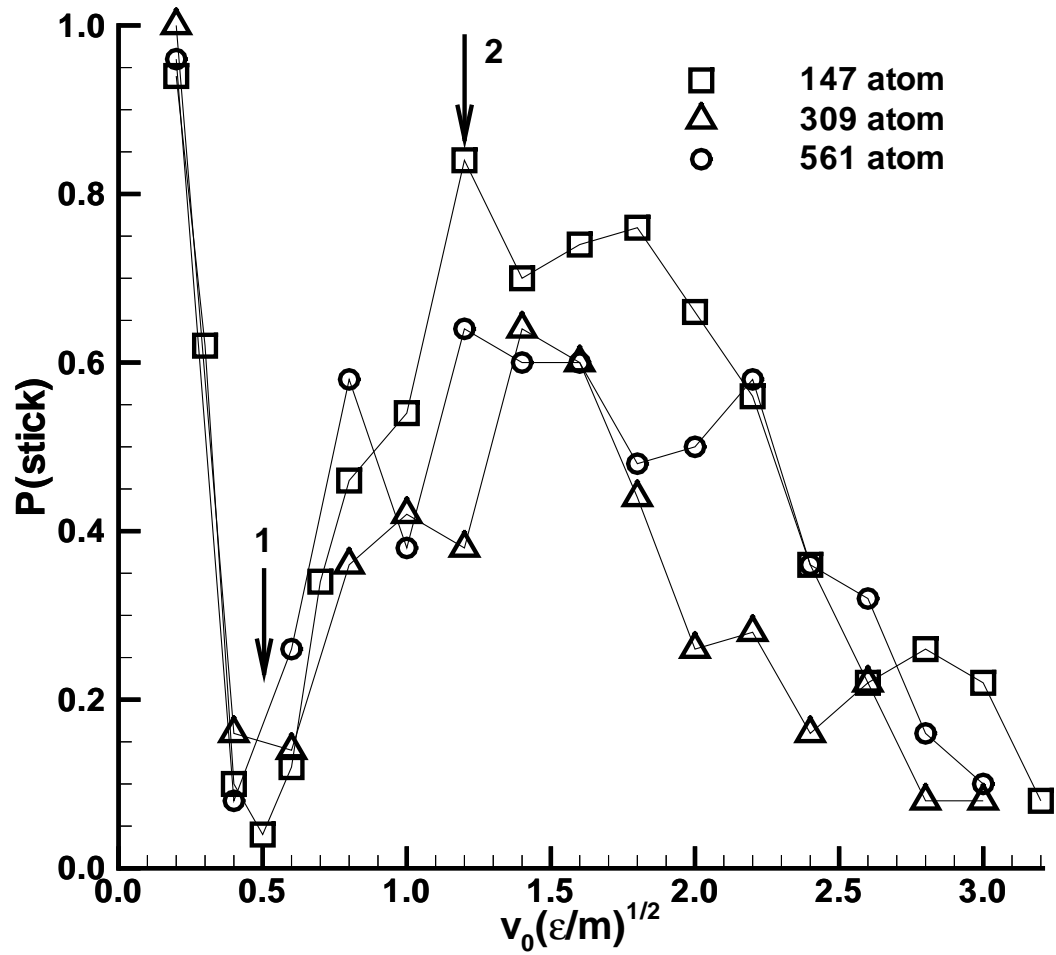}}
\caption{\label{PstickvsV} The probability of sticking as a function of incident velocity for each cluster size. At each velocity we conducted fifty trial simulations with randomly selected cluster orientations. Note the minimum and maximum in adhesion probability (at 1 and 2 respectively).}
\end{figure}

Figure~\ref{PstickvsV} shows the probability of adhesion as a function of velocity. Strikingly, the probability of adhesion is a bimodal function of velocity at all cluster sizes with a second peak in adhesion at $v_0 \sim 1.2 \, (\varepsilon/m)^{1/2}$. At low velocities the cluster always adheres to the surface but at intermediate velocities the cluster starts bouncing. At higher velocities ($0.5 < v_0 < 3.0 \, (\varepsilon/m)^{1/2}$), we see reentrant adhesion followed by reflection in the limit where the velocity approaches the fragmentation regime. We find that this bimodality is also evident at $c=0.3$ and 0.4, but disappears above $c=0.5$, where cluster collisions almost always result in adhesion (at least up to fragmentation), and below $c=0.2$, where only very slow collisions lead to adhesion.     

\begin{figure}
\resizebox{\columnwidth}{!}{\includegraphics{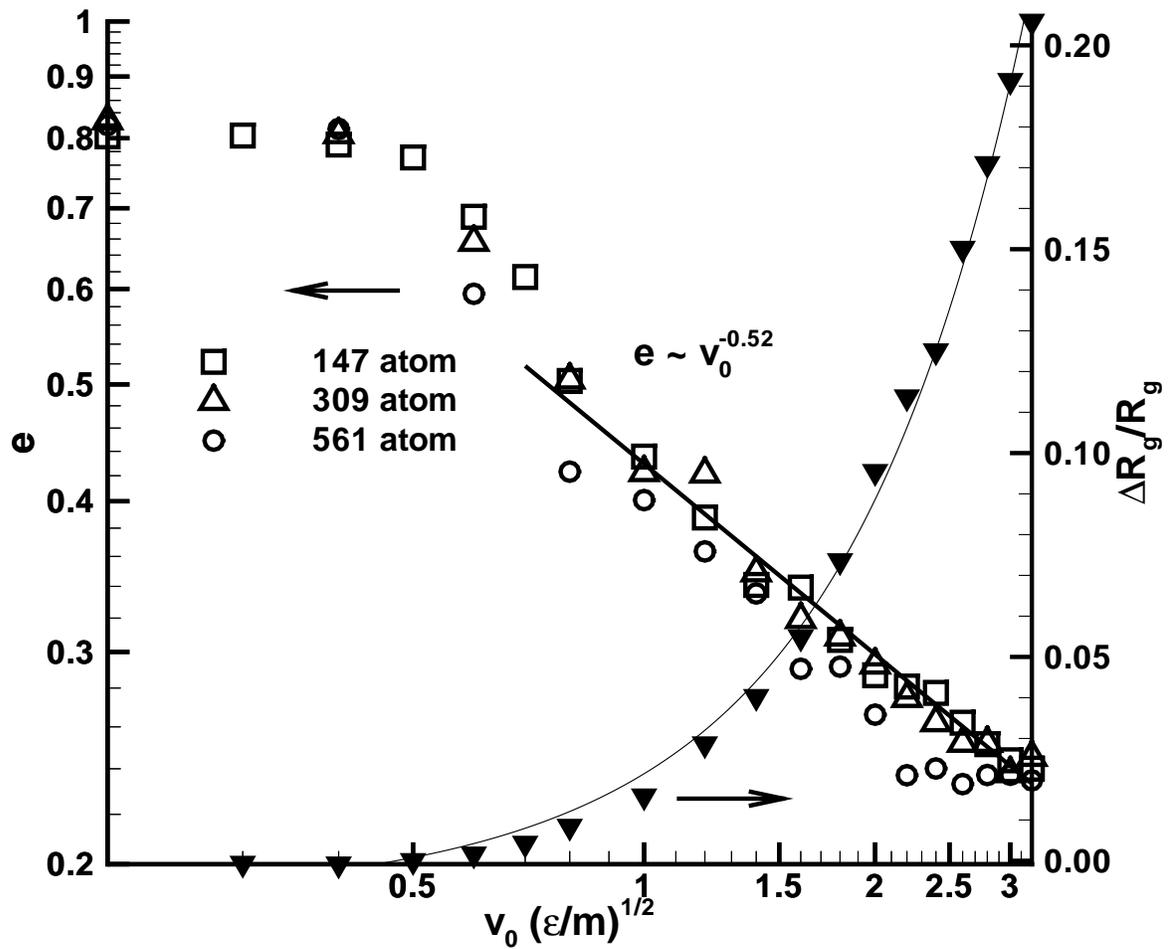}}
\caption{\label{rest}The coefficient of restitution as a function of initial velocity $v_0$ for $c=0.35$ for all three cluster sizes. Also shown (on the right hand axis) is the relative deformation of the 147-atom cluster as a function of velocity which is well-fitted by a quadratic in $v_0$.} 
\end{figure}

The average coefficient of restitution for each cluster size is shown as a function of initial velocity $v_0$ in Fig.~\ref{rest}. Here we define the the coefficient of restitution as the ratio of the maximum velocity after the collision, $v_f$, to the maximum velocity prior to the collision, $v_i$ (see Fig.~\ref{DefnRest}). Note that $e$ is a relatively weak function of cluster size. At low velocities $e$ is approximately constant but at high velocities, $v_0 > 0.7 \,(\varepsilon/m)^{1/2}$, the coefficient exhibits a strong dependence on velocity: $e \sim v_0^{-0.52}$. This is a much stronger dependence on velocity than that given by Hertzian contact mechanics \cite{Johnson87} but is consistent with finite-element simulations of strongly plastic collisions with no adhesion \cite{Wu05}. We note that this dependence on $v_0$ weakens in our simulations as $c$ increases.

Figure~\ref{rest} also shows the relative change in radius of gyration, $\Delta R_{g} / R_g$, of the 147-atom cluster at the moment of peak reflection velocity, $v_f$, as a function of $v_0$. The fit shown is a quadratic in $v_0$ i.e. $\Delta R_{g} / R_g \sim v_0^2$ for $v_0 > 0.5 \, (\varepsilon/m)^{1/2}$. Thus the relative spreading of the cluster during the impact on the surface, $\Delta R_{g} / R_g$, is proportional to the incident kinetic energy. We note that this proportionality is consistent with strong plastic deformation, where the kinetic energy is dissipated largely at the cluster yield stress, $Y$, so that the plastic work $\sim Y \Delta (R^3) \sim Y R^2 \Delta R$ is proportional to the translational kinetic energy $\sim v_0^2 R^3$. We have found that this relationship is relatively insensitive to both the values of c and the cluster sizes examined here. 

\begin{figure}
\resizebox{\columnwidth}{!}{\includegraphics{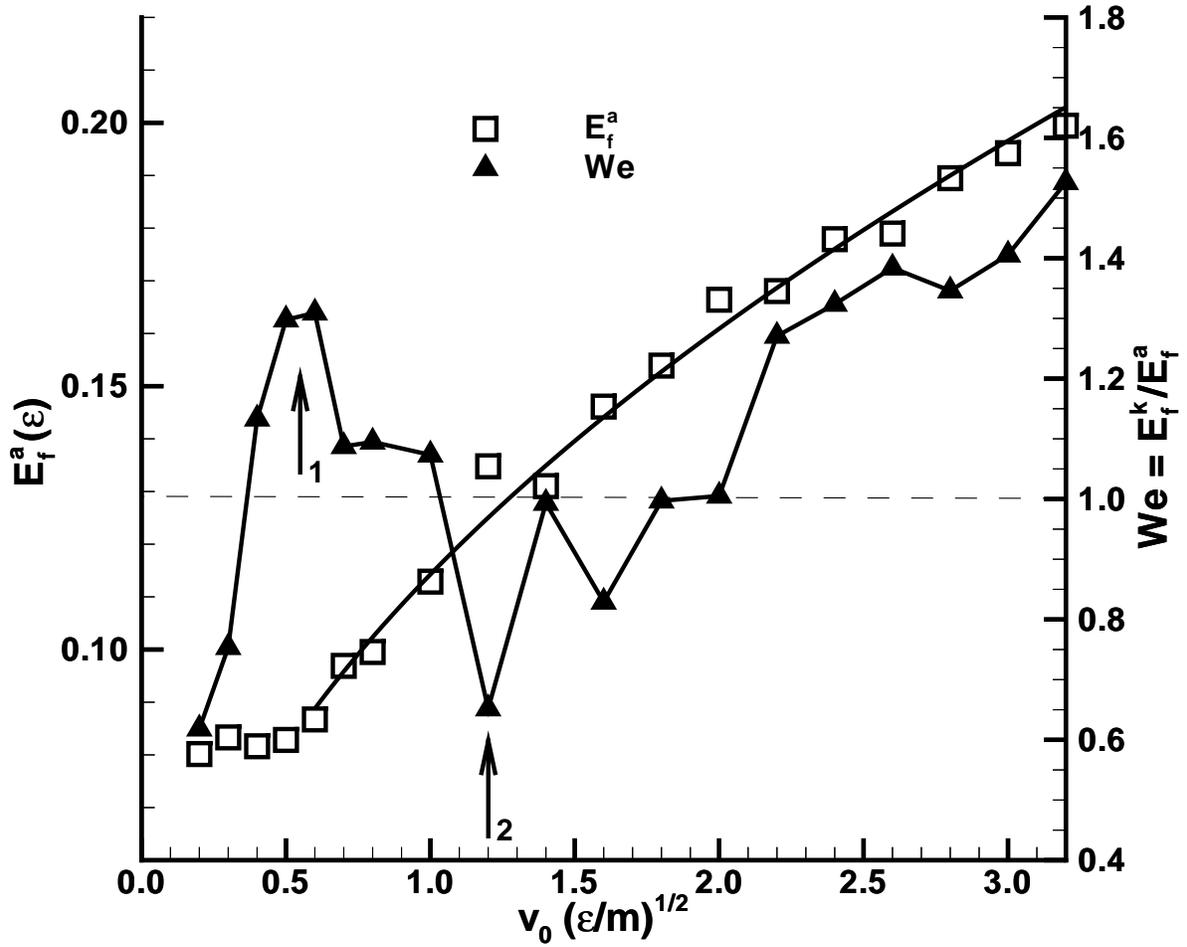}}
\caption{\label{EavsV} The adhesion energy $E^a$ at the peak reflected velocity and the corresponding Weber number as a function of $v_0$ for the 147-atom cluster with $c=0.35$. The fit to the $E^a$ points is a linear function of $v_0^{0.50}$ and is shown by the solid line. Note the maximum (at 1) and minimum (at 2) in the Weber number which correspond respectively to the minimum and maximum in the adhesion probability in Fig.~\ref{PstickvsV}.} 
\thispagestyle{empty}
\end{figure}

The amount of deformation affects the adhesion energy of the cluster by altering the contact area of the cluster with the surface. Fig.~\ref{EavsV} shows the dependence of the adhesion energy $E^{a}_f=-E^{cs}_f$ at the moment of peak reflected velocity $v_f$ for the 147-atom cluster. For $v_0 < 0.5 \, (\varepsilon/m)^{1/2}$, with little deformation of the cluster during the collision, $E^{a}_f$ is constant. At intermediate to large velocities, as the cluster begins to deform on impact, this energy depends on velocity as $v_0^{0.50}$. 

Also shown in Fig.~\ref{EavsV} is the Weber number, $We = E^{k}_f/E^{a}_f$, at the moment of peak reflected velocity. As the average value of $We$ approaches and then exceeds 1, the number of clusters being reflected dramatically increases. In the low deformation regime ($v_0 < 0.5 \, (\varepsilon/m)^{1/2}$), both the coefficient of restitution and the adhesion energy are approximately constant, so $We \sim v_0^2$. This is essentially equivalent to the liquid droplet model\cite{hartley58} used in Ref~\cite{partridge04} to explain the transition from adhesion to reflection in their experiments. However, our estimate of the velocity at which strong plastic deformation occurs\cite{Wu03}, $v^{\dagger} \sim 0.1 \left(Y/ \rho \right)^{1/2} \sim 5-20$ ms$^{-1}$, is exceeded by the estimated cluster impact velocities of 100-200 ms$^{-1}$ in Ref.~\cite{partridge04}. Thus it is probable that the collisions in Ref.~\cite{partridge04} occur in the large deformation (plastic) regime. We note that the metallic clusters used in the experiments are likely to deform plastically at lower energies (relative to their binding energies) than the clusters described by the pair potential here \cite{Holian91}.

At velocities $v_0 > 0.5 \, (\varepsilon/m)^{1/2}$, both the coefficient of restitution and the adhesion energy become dependent on velocity as the cluster undergoes substantial plastic deformation. In particular, the reflected kinetic energy $E^K_f$ goes as $e^2 v_0^2 \sim v_0^{0.96}$ for $v_0 > 0.7 \, (\varepsilon/m)^{1/2}$ and the adhesion energy $E^{a}$ goes as $v_0^{0.50}$ for \, $v_0 > 0.5 \,(\varepsilon/m)^{1/2}$ (as shown in Fig.~\ref{EavsV}). At the onset of this plastic deformation regime, $E^{a}$ grows faster than $E^K_f$, leading to a decrease in $We$ and an increase in adhesion probability. Put simply, the increase in adhesion for $v_0 > 0.5 \,(\varepsilon/m)^{1/2}$ occurs because the clusters are highly deformed (they "pancake"), and so there is an increase in contact area and (consequently) adhesion energy. However at velocities $v_0 > 1.2 \,(\varepsilon/m)^{1/2}$, we find that $We$ increases as $\sim v_0^{0.46}$ as $E^K_f$ begins to dominate, leading to a corresponding decrease in adhesion probability. 

To summarise, we have identified a bimodality in the adhesion probability for clusters incident on a surface with poor adhesive properties. The re-entrant transition seen here is likely to be an example of a more general phenomenon which will occur when ever the Weber number at the transition between the low and strong deformation regimes is near one. Indeed as the Weber number depends on size, we expect that for a given cluster-substrate system there will be a range of cluster sizes where this re-entrant transition may be observed. For example, in metallic clusters, where the onset of strong deformation occurs at lower collision energies, we would expect the reentrant transition to be shifted to larger cluster sizes.

We also find that the dependence of the coefficient of restitution on velocity is much stronger than that predicted by the classical theory of solid-solid collisions \cite{Johnson87} (leading to a strong dependence of the adhesion energy on velocity) although it appears to be consistent with finite-element simulations of strongly plastic collisions in larger particles \cite{Wu05}. This work suggests the possibility of experimental tests of the theory of collisions in nanoscale systems, in addition to the existence of new deposition regimes that could be exploited for device manufacture \cite{partridge04}. Further work will focus on the effects of cluster orientation and non-normal incidence on the cluster reflection.


\clearpage

\clearpage
\thispagestyle{empty}

\clearpage
\thispagestyle{empty}

\clearpage
\thispagestyle{empty}

\clearpage
\thispagestyle{empty}

\clearpage
\thispagestyle{empty}

\end{document}